\documentclass[prb,grupeaddress,twocolumn,aps,showpacs]{revtex4}
\usepackage{epsfig}
\usepackage{dcolumn}
\usepackage{bm}
\usepackage[T1]{fontenc}
\usepackage{graphicx}                    

\begin{document}

\title{Effect of bulk inversion asymmetry on the Datta-Das transistor}

\author{A. {\L}usakowski}
\author{J. Wr\'obel}
\author{T. Dietl\footnote{Address in 2003: Institute of Experimental and Applied Physics, Regensburg University; supported by Alexander von Humbold Foundation}}

\affiliation{Institute of Physics, Polish Academy of Sciences, al. Lotnik\'{o}w 32/46, PL 02-668 Warszawa, Poland}

\date{\today}

\begin{abstract}
A model of the Datta-Das spin field-effect transistor is
presented which, in addition to the Rashba interaction, takes into
account the influence of bulk inversion asymmetry of
zinc-blende semiconductors. In the presence of bulk
inversion asymmetry, the conductance is found to depend significantly on the crystallographic
orientation of the channel. We determine the channel direction optimal
for the observation of the Datta-Das effect in GaAs and InAs-based devices. 
\end{abstract}


\maketitle
Various spin-orbit (SO) effects which are always present
in semiconductor quantum structures, are recognized as an important
issue in the emerging field of spintronic devices. Already the
first proposal of a spin transistor by Datta and Das \cite{Datt90}
relied on the SO-induced splitting of the carrier bands in high
electron mobility transistors (HEMT). As postulated by Bychkov and
Rashba,\cite{Bych84} in such systems, owing to the SO interaction
and interfacial electric field ${\bm E}$ (the structure inversion
asymmetry - SIA), the spins of the electrons flowing with the
velocity ${\bm v}$ precess in a magnetic field ${\bm B}_R \sim {\bm
E}\times {\bm v}$. To take advantage of this fact, Datta and Das
considered a hybrid structure consisting of a ballistic
quasi-one-dimensional (1D) wire patterned of a 2D HEMT  and of 
ferromagnetic source and drain ohmic contacts, serving as spin
injector and detector, respectively. The contacts were assumed to
be magnetized along the current direction and to have 100\% spin
injection/detection efficiency. The calculated conductance for such
a device was found to oscillate as a function of the SO coupling
strength $\alpha$, whose magnitude could be controlled {\it via} a
metal gate located on the top of the structure. Datta and Das
showed that in the lowest order in $\alpha$ the oscillations
depended on neither 1D sub-band index $n$ nor electron wave vector
$k$.

Since the original proposal of Datta and Das, the operation of
their spin field-effect transistor (spin-FET) has been analyzed in
much detail. For example, calculations of device conductance not
only for small but also for large $\alpha$ were carried
out\cite{Mire01} as well as possible effects of electron-electron
interactions were analyzed.\cite{Haus01} Furthermore, an important
problem of spin injection across the ferromagnet/semiconductor
interface was considered from various
viewpoints.\cite{Mats02,Dede02,Meie02} However, in addition to
the SIA (Rashba) term,  owing to
the lack of inversion symmetry in bulk materials -- there exists
the so-called bulk inversion asymmetry (BIA) or Dresselhaus
term.\cite{Dres55} It is known that the combined effect
of both terms modifies considerably spin properties of
zinc-blende heterostructures.\cite{Ohno99Aver99} Therefore, the SIA
and BIA contributions should be considered on equal footing when
analyzing the performance of spin devices.\cite{Schl03} 

In this work, we consider a ballistic quantum wire patterned of a zinc-blende
heterostructure grown along the [001] direction. The Hamiltonian
describing the electron motion takes into account both SIA and BIA
terms. By a numerical solution of the Schr\"odinger equation for
GaAs- and InAs-based systems, we demonstrate the importance, of hitherto
neglected, non-linear terms in the BIA hamiltonian. Moreover, we show that the device
performance depends strongly on the wire direction in respect to
the crystal axes. This allows us to determine channel
orientation corresponding to the largest transconductance for both GaAs and
InAs based devices.

In order to model the Datta-Das transistor, we consider an infinite
quantum wire that extends in the $x$ direction, which is oriented at an angle
of $\phi$ to the [100] crystal axis. The potential confining
electrons in $y$ direction, $V(y)$, is assumed to be parabolic and
characterized by the energy difference
between consecutive harmonic oscillator levels
$\hbar\omega_0$=1 meV in GaAs and $\hbar\omega_0$=3 meV in
InAs, respectively. In order to simulate the
presence of ferromagnetic electrodes, we assume that in the regions
$x < 0$ and $x > L$ the electron
spins experience the magnetic field $|B_{\mbox{\tiny out}}| =
10^3$~T, which is strong enough to ensure the entire spin
polarization. Following Datta and Das we take the vectors ${\bm
B}_{\mbox{\tiny out}}$ to point in the $x$ direction and consider
the case of their parallel or anti-parallel relative orientations.
In the channel region, $0 < x < L$, $B_{\mbox{\tiny out}}$ decays
adiabatically to zero. 
For $x \parallel [100]$, the SO Hamiltonian is given by\cite{Piku95}
\begin{eqnarray}
\label{eq1} H_{SO}=
 i\alpha \left(\sigma_y \partial_x -\sigma_x\partial_y\right) \nonumber\\
+i\gamma\left(\sigma_x \partial_x\left(\partial_y^2+
\langle k_z^2\rangle\right)-\sigma_y
\partial_y\left(\langle k_z^2\rangle-\partial_x^2\right)\right), 
\end{eqnarray}
where $\sigma_x$ and $\sigma_y$ are the Pauli matrices. The terms
proportional to $\alpha$ and $\gamma$ corresponds to the SIA and BIA,
respectively, whose effect on the band splitting has been thoroughly 
examined.\cite{Pfef99}  Here, we estimate the magnitude of $\alpha$ 
and $\langle k_z^2\rangle$ according to Ref.~\onlinecite{Piku95}
\begin{eqnarray}
\label{alpha}
\alpha=\frac{\hbar^2}{2m}\frac{\Delta\left(2E_g+\Delta\right)}
{E_g\left(E_g+\Delta\right)\left(3E_g+2\Delta\right)} \frac{2\pi
e^2 N_s}{\kappa};\\
\label{gamma}
\langle k_z^2 \rangle = \frac{1}{4}(16.5 \pi e^2 m N_s/\kappa \hbar^2)^{1/3},
\end{eqnarray}
which gives values comparable with those obtained from the more
refined theories, at least in low Fermi energy limit. 
In Eq.~(\ref{alpha}) $\kappa$ is
the static dielectric constant, $m$ is the effective mass,  $E_g$
is the energy gap, and $\Delta$ is the spin-orbit splitting in the
valence band.\cite{uwaga3} For example, for typical areal electron
density $N_s = 5\times 10^{11}$~cm$^{-2}$, this approach implies
$\alpha = $1.7 meV\AA \ and 42.0 meV\AA \ for GaAs- and InAs-based
heterostructures, respectively. The BIA parameter $\gamma$ is equal to
25 eV\AA$^3$ for GaAs and 130 eV\AA$^3$ for InAs.\cite{Piku95,Silv94}

By transforming the Hamiltonian
(\ref{eq1}) to a new coordinate system such that the angle between
new and old $x$ axes is $\phi$ we obtain,
\begin{eqnarray}
\label{eq2} H_{SO}=
 i\alpha \left(\sigma_y \partial_x -\sigma_x\partial_y\right) \nonumber \\
-i\gamma\left[\sigma_y\sin(2\phi)\left(\frac{1}{2}\partial_x^2 
-\frac{1}{2}\partial_y^2+\langle k_z^2\rangle \right) \right . \nonumber \\
\left . 
\vphantom{\frac{1}{2}}
- \sigma_x\cos(2\phi)\left(\partial_y^2 +\langle k_z^2\rangle\right)\right]
\partial_x \nonumber \\ 
-i\gamma\left[
\vphantom{\frac{1}{2}}
\sigma_y\cos(2\phi)\left(\partial_x^2 + \langle k_z^2\rangle \right)
\right . \nonumber \\
+ \left . \sigma_x\sin(2\phi)\left(\frac{1}{2}\partial_y^2 -
\frac{1}{2}\partial_x^2 +\langle k_z^2\rangle \right)\right]\partial_y. 
\end{eqnarray}
We see that only the Dresselhaus (BIA) term changes, while the
Rashba (SIA) term remains invariant with respect to the rotations
around the $z$ axis.\\
The full Schr\"odinger equation for our calculations reads,
\begin{eqnarray}
\label{eq23}
\left(-\frac{\hbar^2}{2m}\left(\partial_x^2+\partial_y^2\right) + V(y)
+ H_{SO} \right . \nonumber \\
\left . + \frac{1}{2}g\mu_B
{\bm\sigma}\cdot {\bm B}_{\mbox{\tiny out}}(x)\right) \Psi(x,y)=E_F\Psi(x,y),
\end{eqnarray}
where $E_F$ is the Fermi energy.

To calculate conductance in the non-interacting electron model we
employ the recursive Green function method\cite{Lee81MacK83Ciep91} in
a form developed in Ref.~\onlinecite{Zozo96}. 
We expand the spinor wave function $\Psi(x,y)$ into a series
\begin{equation}
\label{series}
\Psi(x,y)=\sum_{n} \chi_n(x) \psi_n(y),
\end{equation}
where $\psi_n(y)$ are the eigenfunctions of one dimensional harmonic
oscillator potential
\begin{equation}
\label{harmonic}
\left(-\frac{\hbar^2}{2m}\partial_y^2 +
V(y)\right)\psi_n(y)=\epsilon_n \psi_n(y).
\end{equation}
By using the orthonormality property of these functions, it is easy to
derive a set of differential equations for $\chi_n(x)$. 

The problem we encounter is the appearance of third order derivative with respect
to $x$ in the term $-i\gamma \sin(2\phi) \partial_x^3 \chi_n(x)$, which
 originates from the BIA part of the Hamiltonian. We replace
 this term by 
$i\gamma\sin(2\phi)k_x^2(n)\partial_x \chi_n$ where $k_x^2(n)\equiv
\frac{2m}{\hbar^2}\left(E_F-\epsilon_n\right)$ is the wave vector for
$n$'th mode in the absence of spin-orbit coupling. This approximation
is justified by the fact that for typical experimental situations, 
$\gamma k_F^3 << \frac{\hbar^2k_F^2}{2m}$. Indeed, for GaAs for $k_F
\sim 10^{8}{\rm m}^{-1}$ and $\gamma = 25$ meV\AA$^3$, 
$\gamma k_F^3 \sim 4.0 \times {\rm 10}^{-24}$~J, which is much smaller
than $\frac{\hbar^2 k_F^2}{2m} \sim 8\times {\rm 10}^{-22}$~J. 
In other words, the spatial changes of the wave function are
governed mainly by the $-\frac{\hbar^2}{2m}\partial_x^2$ term in the
Hamiltonian. Treating $\chi_n$ as the components of a vector ${\bm \chi}$ 
we arrive to
\begin{equation}
\label{eq100}
-\frac{\hbar^2}{2m}R\left(\partial_x^2 {\bm \chi} + 2i Q \partial_x {\bm \chi}
\right) + \left(h-E_F\right){\bm \chi}=0,
\end{equation}
where matrix $Q \equiv R^{-1}P$ and 
\begin{equation}
\label{eq101}
R_{nn'}=\delta_{nn'} + i\frac{2m}{\hbar^2}\gamma  \left[ \sigma_y\cos(2\phi) - \frac{1}{2} \sigma_x
\sin(2\phi) \right] D^{(1)}_{nn'};
\end{equation}
\begin{eqnarray}
\label{eq102}
P_{nn'}=\frac{m}{\hbar^2}\gamma\{\delta_{nn'}\sigma_y\sin(2\phi) [\langle k_z^2 \rangle
-\frac{1}{2} k_x^2(n)]  \nonumber \\
 - \delta_{nn'} \langle k_z^2 \rangle \sigma_x\cos(2\phi)
-[\sigma_x\cos(2\phi)  \nonumber \\
 +\frac{1}{2}\sigma_y\sin(2\phi)]
D^{(2)}_{nn'} \} -\frac{m}{\hbar^2}\alpha \delta_{nn'}\sigma_y;
\end{eqnarray}
\begin{eqnarray}
\label{eq103}
h_{nn'}=\left(\epsilon_n+ \frac{1}{2}g\mu_B
{\bm\sigma}\cdot {\bm B}_{\mbox{\tiny out}}(x)\right)\delta_{nn'}
\nonumber \\ 
-i\gamma \left[\langle k_z^2 \rangle \left(\sigma_y\cos(2\phi) +
\sigma_x \sin(2\phi)\right) D^{(1)}_{nn'}\right . \nonumber \\
\left .+\frac{1}{2}\sigma_x\sin(2\phi) D^{(3)}_{nn'}\right]-i\alpha \sigma_x D^{(1)}_{nn'},
\end{eqnarray}
and the the matrix $D^{(n)}_{mn}=\langle \chi_m|\partial_y^{(n)}|\chi_n\rangle$.
After discretization along $x$ direction (with the lattice constant
$a$) we finally obtain 
\begin{eqnarray}
\label{eq104}
-tV_{x,x+a}{\bm \chi}(x+a) - tU_{x,x-a}{\bm \chi}(x-a) + 2tR{\bm \chi}(x) \nonumber \\
-\frac{\hbar^2}{2m}RQ^2{\bm \chi}(x) +\left(h-E_F\right){\bm \chi}(x)=0,
\end{eqnarray}
where $t=\frac{\hbar^2}{2ma^2}$, $V_{x,x+a}=Re^{iaQ}$,
$U_{x,x-a}=Re^{-iaQ}$. This form is appropriate for the 
conductance calculation.

\begin{figure}
\includegraphics[scale=0.3,angle=0]{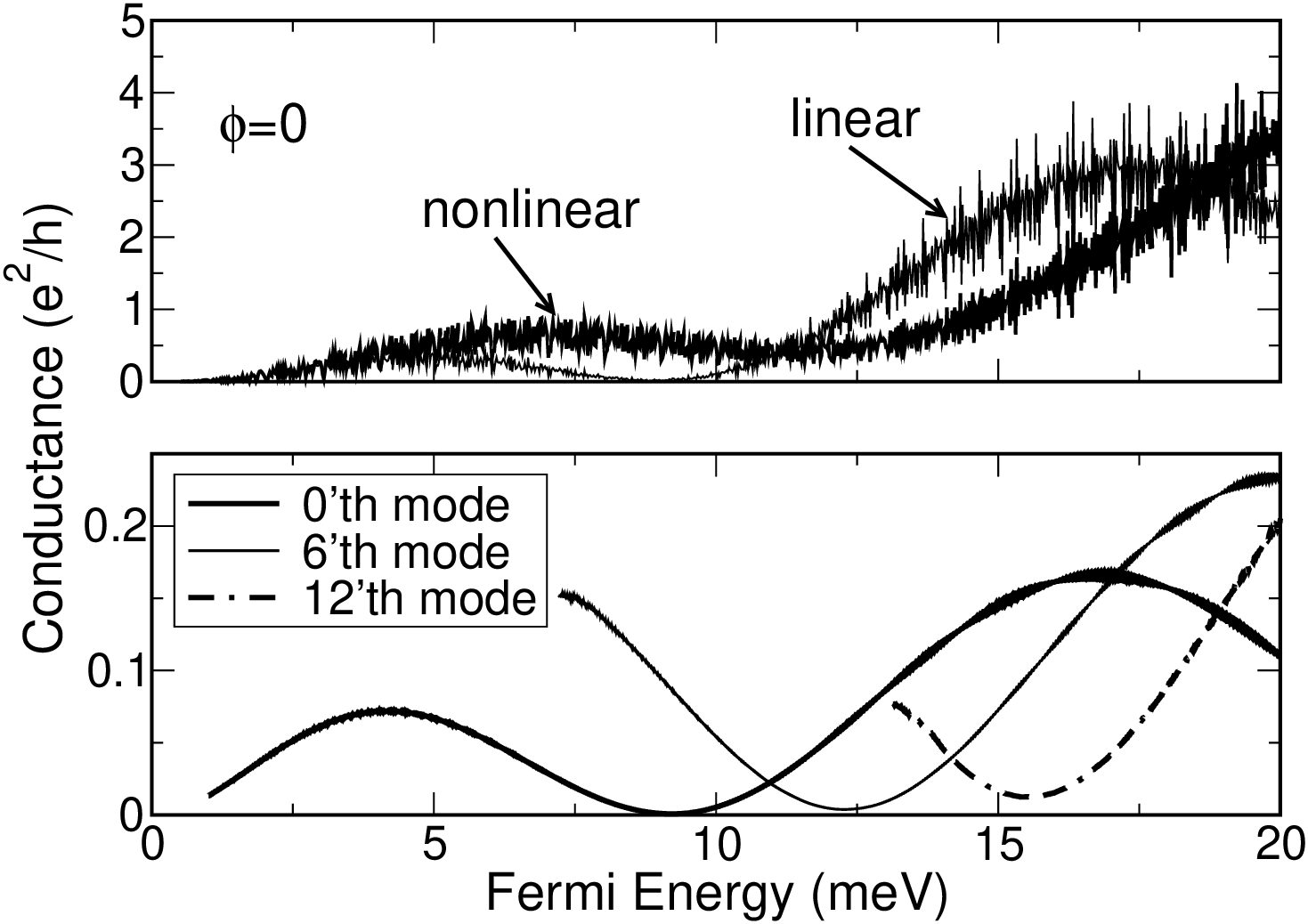}
\caption{Source-drain conductance of a GaAs-based device as a
function of the Fermi energy for $\phi$=0 ([100] channel direction) and anti-parallel electrode
magnetizations. Upper panel - comparison of linear and nonlinear
cases. Lower panel - partial contributions of three chosen modes to the
conductance.} \label{fig1}
\end{figure}

Before presenting our results we discuss briefly the "linear"
approximation to the BIA term. Within this approximation second order derivatives
$\partial_x^2$ and $\partial_y^2$ are neglected
in Eq.~(\ref{eq2}). 
On the basis of this linear approximation one can easily understand
the influence of the channel crystallographic orientation on
the performance of spin-FET. Let us consider a transversal running
mode characterized by wave vector $k$. Following Datta 
and Das \cite{Datt90} we neglect intermode coupling 
caused by spin-orbit terms. Then, it is possible to define Rashba and
Dresselhaus "magnetic fields" ${\bm B}_R=\alpha k(0,-1)$ and ${\bm B}_D=\gamma
\langle k_z^2\rangle 
k(-\cos 2\phi, \sin 2\phi)$. Their sum ${\bm B}_{SO}={\bm B}_R+{\bm B}_D$
determines the spin precession frequency. Importantly,
the magnitude and orientation of ${\bm B}_{SO}$ with respect
to the $x$ axis depend on the angle $\phi$. That
is why the precession frequency varies with the channel
crystallographic orientation. It is 
easy to find out that for $\phi=135^{\circ}$, ${\bm
B}_{SO}$ is perpendicular to the $x$ axis, and its length assumes a maximum value. 
Thus, for the $[\bar{1}10]$ channel orientation, in the linear approximation, the period of
the oscillations is the shortest, and thus the Datta-Das effect best visible. 
These qualitative considerations have been
confirmed by our numerical calculations performed in the linear
approximation. However, in neither GaAs nor InAs the linear approximation is justified for
the experimentally relevant Fermi energy range.

\begin{figure}
\includegraphics[scale=0.3,angle=0]{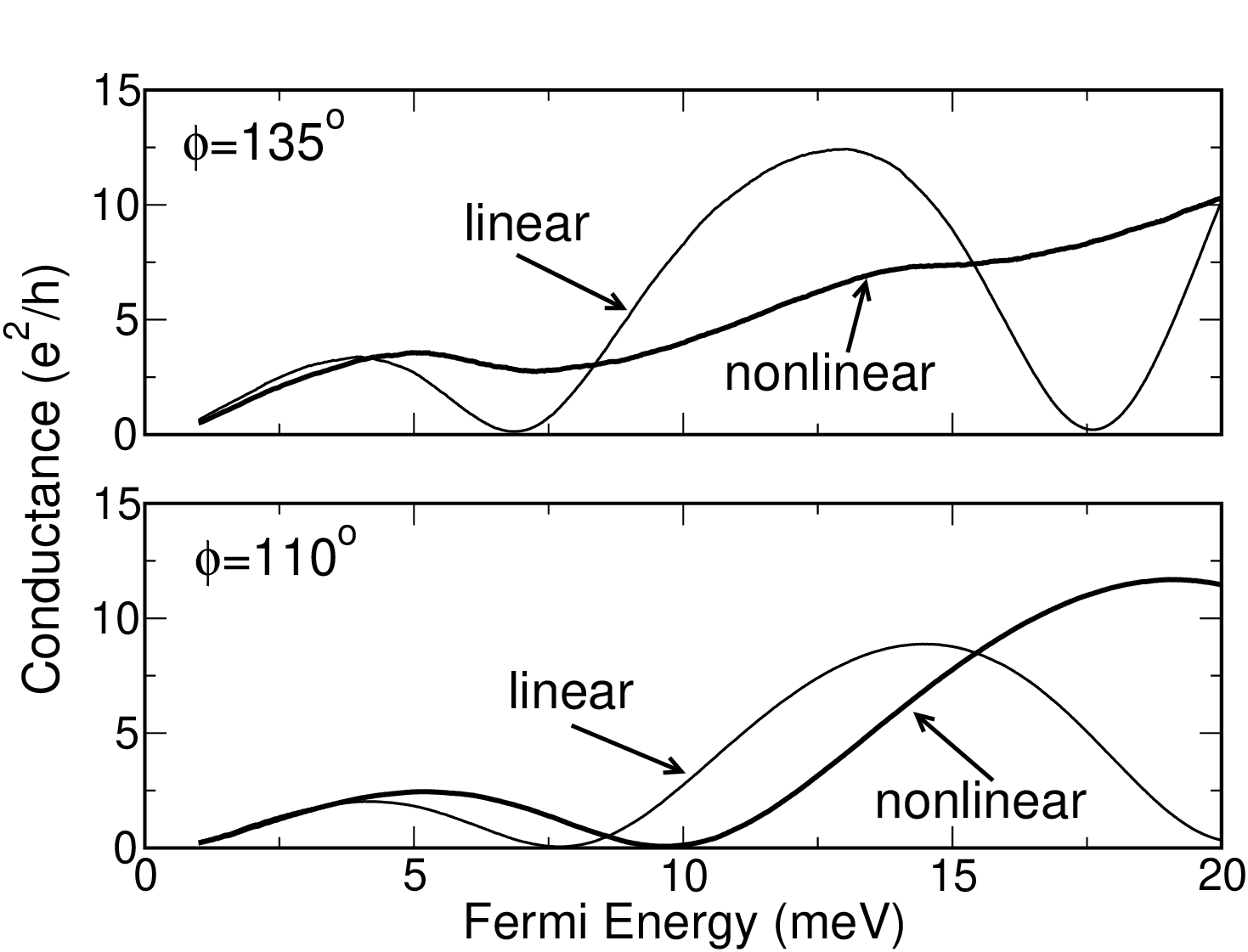}
\caption{Source-drain conductance of a GaAs-based device as a
function of the Fermi energy for $\phi=135^{\circ}$ (upper panel)  and
for for $\phi=110^{\circ}$ (lower panel). } \label{fig2}
\end{figure}

We start the discussion of the full nonlinear model from results for the device
oriented along the [100] direction ($\phi=0$). As shown in the upper panel of Fig.~1, 
for a given relative magnetization of the contacts, two types of conductance oscillations 
as a function of the Fermi energy appear: the long-period Datta-Das oscillations and the
short-period Fabry-Perot oscillations resulting from interference
between transmitted and reflected spin wave functions by the two 
contacts.\cite{Mats02} In real experiments, the latter are expected to
be washed out by thermal broadening and effects of non perfect geometry. 
Therefore, in the following pictures we show 
smoothed data obtained by averaging with the Fermi energy window of 1.0 meV.  
We realize that in the nonlinear case, contributions to the conductance originating from particular modes
oscillate with  different periods.  When the Fermi energy or, equivalently,
electron concentration grows, every 1 meV a new mode, with its own period, starts to
participate in the transport. We illustrate this effect in the lower 
panel of Fig.~\ref{fig1}, where a partial conductance due
to 0th, 6th and 12th mode is depicted. This leads to the disappearance 
of the conductance minima 
when the nonlinear terms in the SO Hamiltonian are taken into
account, as shown in Fig.~\ref{fig2} (upper panel). By analyzing the conductance 
for different angles  $\phi$ 
we have found that the optimal angle for the
observation of Datta-Das effect in GaAs is $\phi=110^{\circ}$ (see
Fig.~\ref{fig2} lower panel). 

\begin{figure}
\includegraphics[scale=0.3,angle=0]{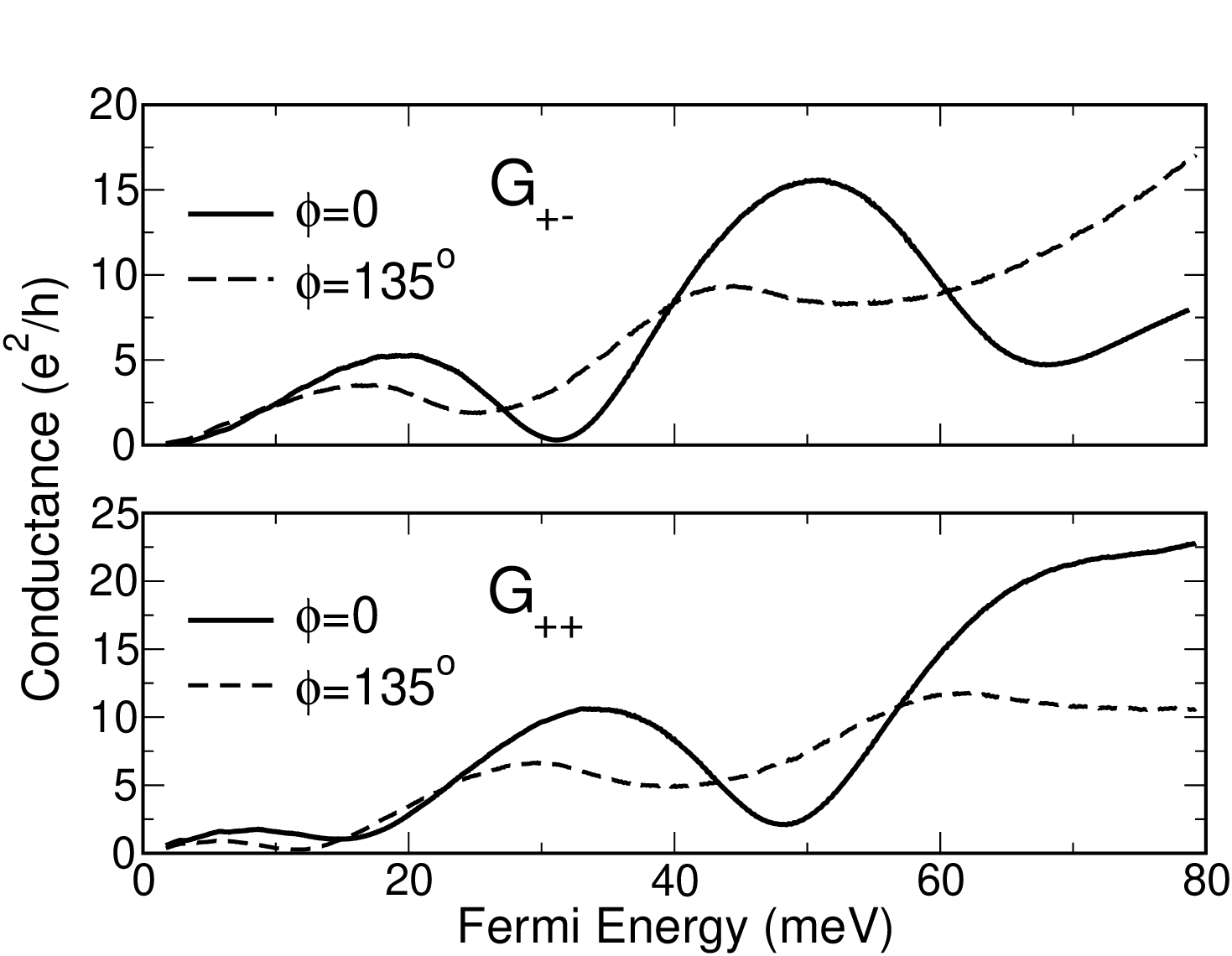}
\caption{Source-drain conductance of a InAs-based device as a
function of the Fermi energy for $\phi=0^{\circ}$ and
$\phi=135^{\circ}$ for the opposite (upper panel) and
parallel (lower panel) source-drain polarizations. } \label{fig3}
\end{figure}

Turning to the InAs-based transistors, we note that different ratios 
of parameters describing SIA and
BIA terms result in a different behavior of the
conductance as a function of the transistor direction. For InAs the
spin-orbit parameters are about order of  
magnitude larger, however, the electron mobility in InAs is low
comparing to GaAs so the length of the quantum wire have to be shorter,
otherwise we cannot assume ballistic transport regime. In calculations
we have assumed $L$=4.0 $\mu$m.\cite{Erom02} The results for the optimal angle
$\phi$=0$^{\circ}$  are presented in Fig.~\ref{fig3}. They are compared
to the case $\phi$=135$^{\circ}$ which is the optimal angle determined
within the linear approach. In Fig.~\ref{fig3} we show conductance
for opposite (upper panel) and parallel (lower panel) source-drain
polarizations, respectively. 

\begin{figure}
\includegraphics[scale=0.3,angle=0]{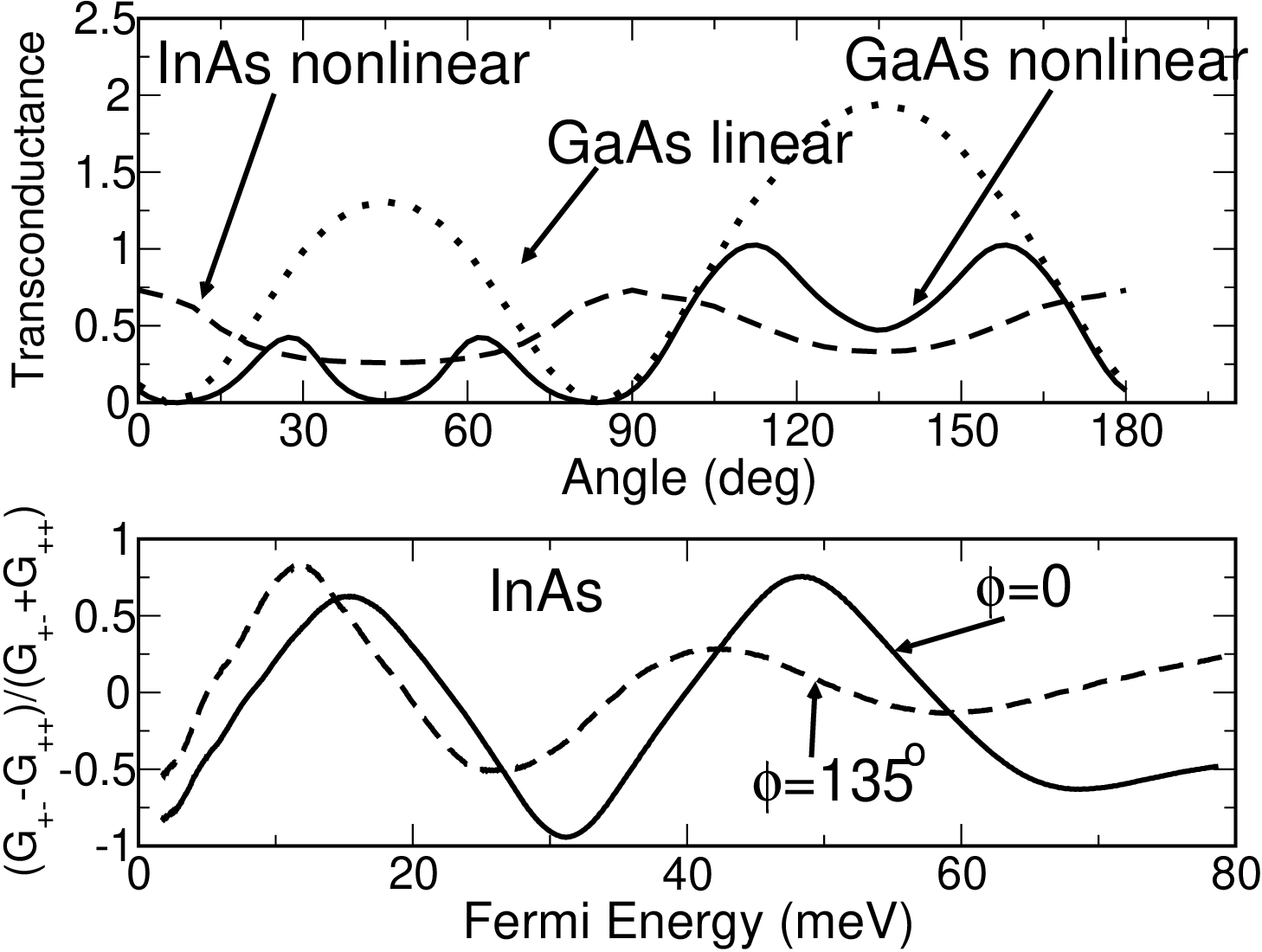}
\caption{Upper panel - Transconductance as a function of channel's
direction. Lower panel - Relative difference of conductances for
opposite and parallel source-drain polarizations. } \label{fig4}
\end{figure}

The angle dependence of conductance for GaAs and InAs devices 
is summarized  
in the upper panel of Fig.~\ref{fig4}, where we plot the transconductance,
$|dG/dE_F|$,  a quantity of particular experimental interest. 
The transconductance in Fig.~\ref{fig4} is defined as the absolute value
the derivative of conductance with respect to the Fermi energy. For a given angle
$\phi$, we take into account the energy
region, where the conductance decreases. Indeed, a decrease of $G$ is
directly connected with the Datta-Das effect while
its increases is connected also 
with the increasing number of modes participating in the transport. 
In the lower panel of Fig.~\ref{fig4} we present  
the Fermi energy dependence of another quantity which is also important 
in the experimental search for the Datta-Das effect,\cite{Mats02,Meie02} 
$P=(G_{+-}-G_{++})/(G_{+-}+G_{++})$, where $G_{++}$ and $G_{+-}$ are
conductances for parallel and opposite source-drain polarizations
respectively. Our results imply, therefore, that there is much
room for the enhancement of the Datta-Das signal by selecting
an appropriate crystallographic orientation of the device.

In conclusion, we have developed a theoretical approach suitable to
numerical studies of ballistic electron transport in semiconductor
wires taking into account spin-orbit effects. The model has been applied
to the Datta-Das spin-FET of zinc-blende semiconductors.  There are two main
results of our model calculations. First, there is a strong dependence of the
transistor conductance on its direction with respect to 
crystallographic axes. Secondly, our calculations explicitly show the
importance of nonlinear terms in BIA Hamiltonian. For particular
model of the spin-orbit parameters, Eq.~(\ref{alpha}) and
Eq.~(\ref{gamma}), we have also 
determined transistor direction optimal for the observation of the
Datta-Das effect, which is $[\bar{1}20]$ and [100] for the GaAs and InSb-based devices, 
respectively. 

The work was supported by the KBN grant PBZ-KBN-044/P03/2001,
FENIKS project (EC: G5RD-CT-2001-0535), and Centre of Excellence
CELDIS (ICA1-CT-2000-70018) within the 5th Framework Programme of
European Commission. The authors thank P.~Pfeffer for 
discussions concerning the role of nonlinear terms in BIA Hamiltonian.

\end{document}